\documentclass[a4paper,11pt]{article}
\usepackage{cite}
\usepackage{jheppub} 
\usepackage{etoolbox}
\patchcmd{\maketitle}{\@fpheader}{}{}{}
\usepackage[T1]{fontenc} 
\usepackage[utf8]{inputenc} 
\usepackage{microtype} 
\usepackage{lmodern} 
\usepackage{mdframed} 
\usepackage{mathtools}
\usepackage{graphicx}
\usepackage{cancel} 
\usepackage{todonotes} 
\usepackage[normalem]{ulem} 
\usepackage{soul} 
\usepackage{amsmath}
\usepackage{amsfonts}
\usepackage{amssymb}
\usepackage{stmaryrd}
\usepackage[mathscr]{euscript}
\usepackage{mathtools}
\usepackage{mathrsfs}
\usepackage{multicol}
\usepackage{mleftright} \mleftright
\usepackage{tensor} 
\usepackage{braket} 
\usepackage{mathrsfs} 
\usepackage{bbold}
\usepackage{color}
\usepackage{braket}
\usepackage{slashed}
\usepackage{hyperref}

\newcommand{\tJ}{\tilde J}
\newcommand{\tP}{\tilde P}

\newcommand{\tY}{\tilde Y}

\newcommand{\tLambda}{\tilde \Lambda}

\newcommand{\comment}[1]{}

\providecommand{\sc}{{ }}
\renewcommand{\sc}{{ }}

\DeclareMathAlphabet{\mathfs}{U}{rsfs}{m}{n}                     %

\newcommand{\n}{\nonumber}
\newcommand{\be}{\nopagebreak[3]\begin{equation}}
\newcommand{\ee}{\end{equation}}
\newcommand{\bee}{\nopagebreak[3]\begin{equation*}}
\newcommand{\eee}{\end{equation*}}
\newcommand{\ba}{\nopagebreak[3]\begin{eqnarray}}
\newcommand{\ea}{\end{eqnarray}}
\newcommand{\baa}{\nopagebreak[3]\begin{eqnarray*}}
\newcommand{\eaa}{\end{eqnarray*}}
\newcommand{\bal}{\nopagebreak[3]\begin{aligned}}
\newcommand{\eal}{\end{aligned}}

\newcommand{\bseq}{\nopagebreak[3]\begin{subequations}}
\newcommand{\eseq}{\end{subequations}\noindent}

\title{Electric/Magnetic Newton-Hooke and Carroll Jackiw-Teitelboim Gravity}
\author[1]{Luis Avilés}
\author[2]{Joaquim Gomis}
\author[3]{Diego Hidalgo}
\author[4,5]{Jorge Zanelli}

\affiliation[1]{Instituto de Ciencias Exactas y Naturales (ICEN), Universidad Arturo Prat, Avenida Arturo Prat Chacón 2120, 1110939, Iquique, Chile}
\affiliation[2]{Departament de F\'isica Qu\`{a}ntica i Astrof\'isica and Institut de Ci\`{e}ncies del Cosmos (ICCUB), Universitat de Barcelona, Mart\'i i Franqu\`{e}s 1, E-08028 Barcelona, Spain}
\affiliation[3]{Science Institute, University of Iceland, Dunhaga 3, 107 Reykjav\'ik, Iceland}
\affiliation[4]{Centro de Estudios Cient\'ificos (CECs), Av. Arturo Prat 514, Valdivia, Chile}
\affiliation[5]{Universidad San Sebastián, Av. General Lagos 1163, Valdivia, Chile}
\emailAdd{luaviles@unap.cl}
\emailAdd{joaquim.gomis@ub.edu}
\emailAdd{dhidalgo@hi.is}
\emailAdd{jorge.zanelli@uss.cl}
\preprint{{\bf } }

\abstract{We construct the electric and magnetic Newton-Hooke and Carroll Jackiw-Teitelboim gravity theories using the isomorphism of Newton-Hooke$_\pm$ and (A-)dS Carroll algebras in $(1+1)$-spacetime dimensions. The starting point is the non-relativistic and Carroll version of Jackiw-Teitelboim gravity without restrictions on the geometry studied in 
\cite{Gomis:2020wxp}. 

}

\begin{document}
\maketitle
\flushbottom

\section{Introduction}\label{intro}
There has recently been a surge in interest in non-Lorentzian theories and their associated geometries. The reasons are several: their applications in condensed matter physics \cite{SachdevBook, ZaanenBook}, non-relativistic holographic theories \cite{Barnich:2009se,Bagchi:2012cy,Bagchi:2019xfx,Ciambelli:2020eba,Campoleoni:2022wmf}, Einstein gravity and black holes \cite{Blau:2015nee,Penna:2018gfx,Donnay:2019jiz,deBoer:2021jej,Perez:2021abf,Perez:2022jpr,Fuentealba:2022gdx}, and also in hydrodynamics \cite{Ciambelli:2018xat,Ciambelli:2018wre,Freidel:2022bai}. 

Non-Lorentzian theories can arise from the non-relativistic (or Galiliean, {\it i.e.} $c\rightarrow \infty$) and Carroll ($c\rightarrow 0$) limits of Lorentzian theories. In particular, for electromagnetism, Le Bellac and Lévy-Leblond showed that there exist two different Galilean limits: the so-called  ``electric'' and ``magnetic'' limits. Later, the Carroll analogue was derived in \cite{Duval:2014uoa}. Also, it has been recently shown that the Hamiltonian formulation of general relativity admits at least two Carroll limits \cite{Henneaux:2021yzg}. The so-called magnetic case is equivalent to the Carroll theory of gravity, defined through a gauging of the Carroll algebra \cite{Campoleoni:2022ebj}. \\
Alternatively, non-Lorentzian theories can also be built from non-relativistic and Carroll groups using the tools of gauge theory without performing a limit process \cite{Bergshoeff:2017btm}. So far, it is not clear how to identify the “electric” and “magnetic” analogue theories in this context. However, based on the results obtained from the Lagrangian formulation of non-Lorentzian field theories \cite{Bergshoeff:2022eog}, it is possible to notice that an “electric” non-relativistic field theory implies the inclusion of extra fields while the “magnetic” one does not. The opposite is true for the Carroll case; that is, we have a “magnetic” Carroll field theory when extra fields are included in the theory and an “electric” Carroll theory when they are not. Here, the terminology magnetic and electric is based on this fact. \\
The relationship among the non-relativistic and Carroll theories \cite{Barducci:2018wuj,Gomis:2022spp} in a new way has also been studied \cite{axelericq}. The main idea is that given a non-relativistic or Carroll theory, we can construct two more models with additional fields restricting the initial dynamics. This fact will play a very prominent role in this work.
\\ In this note, we further analyze the relation between electric and magnetic gravities in $(1+1)$-dimensions. In other words, we construct the corresponding Jackiw-Teitelboim (JT) gravity theories with extra fields playing the role of Lagrange multipliers. In this case, the Newton-Hooke
algebras and the corresponding Carroll ones are isomorphic. We start by considering the non-relativistic and Carroll BF gravities introduced in \cite{Gomis:2020wxp} and following the procedure of \cite{axelericq} we construct new non-relativistic and Carroll gravity theories with extra fields that further restrict the dynamics. We also build new gravity theories without introducing an $U(1)$ gauge field. 
\section{Basics on non-relativistic JT gravity in the first order formalism}\label{resume}
A first-order formulation of JT gravity can be defined by gauging the (A-)dS$_2$ symmetry and constructing a BF theory \cite{Fukuyama:1985gg,Isler:1989hq,Chamseddine:1989yz}. The gauge algebra is given by $\mathfrak{so}(1,2)$ in the A-dS case ($\tLambda<0$) and $\mathfrak{so}(2,1)$ for dS ($\tLambda >0$). The generators are based on the Lie algebra\footnote{We use the conventions  $\epsilon_{01}=-\epsilon_{10}=1$, and $\eta_{ab}={\rm diag}(-,+)$ is the two-dimensional Minkowski metric.}
\be \label{adS2algebra.}
\left[ \tJ, \tP_a \right] =  {\epsilon_a}^b \tP_b \,,  \hskip.9truecm  \left[ \tP_a, \tP_b \right]  =-\tLambda \, \epsilon_{ab}\, \tJ\,,
\ee
where $\tP_a$ stand for translations, $\tJ$ is the boost generator, $\tLambda $ is the cosmological constant, and $a=0,1$. This algebra admits the following most general non-degenerate invariant bilinear form 
\be\label{ads2invt}
\left\langle \tJ, \tJ \right\rangle =\tilde{\gamma}_{0}  \,, \hskip.8truecm  \left\langle \tP_a, \tP_b \right\rangle =- \tilde{\gamma}_{0}\, \tLambda \, \eta_{ab}\,, 
\ee
where $\tilde{\gamma}_{0} \neq 0$ is an arbitrary constant. In order to define a BF theory we consider a scalar field $B$, valued on the (A-)dS$_2$ algebra
\be\label{JTB}
 B=\Phi^a\, \tP_a + \Phi\, \tJ\,,
\ee
and a set of one-form gauge fields,  $E^a=E^a_\mu\, dx^\mu$ and 
 $\Omega=\Omega_\mu\, dx^\mu$, corresponding to the zweibein form and the dual spin connection $\Omega\equiv- \frac{1}{2}\epsilon_{ab} \, \Omega^{ab}$, respectively. Here the space-time indexes run as $\mu=0,1$. The gauge fields define the (A-)dS$_2$ connection one form
\be\label{JTconnection}
A=E^a\, \tP_a +\Omega\, \tJ\,,
\ee
together with the corresponding curvature two-form $F=dA +\left[A,A\right]$, given by\footnote{Wedge product among differential forms is assumed, for instance $E^a  E^b=E^a  \wedge E^b = -E^b  E^a $.}
\ba
F& =&R^a(\tP)\tP_a+R(\tJ)\tJ\\
R^a(\tP) & =& dE^a - \epsilon^a_{\;\;b} \, \Omega E^b  \\
R(\tJ) & = & d\Omega  -\frac{\tLambda}{2} \epsilon_{ab}E^a E^b \,.
\ea
The (A-)dS BF action then reads
\be \label{JTBFaction}
S[B,A]=\int \left\langle B, F \right\rangle   = \tilde{\gamma}_{0}\int \left(\Phi R(\tJ)-\tLambda \Phi_a R^a(\tP) \right)\,.
\ee
The action \eqref{JTBFaction} is invariant under infinitesimal gauge transformations 
\be
\delta  A  = d  \tilde{\chi}+ [A, \tilde{\chi}]\,,~~~~~~~~~~~\delta B  = [B,\tilde{\chi}]\,,
\ee
where $\tilde{\chi} =  \tilde{\lambda}^a \tP_a + \tilde{\lambda} \tJ$ is a gauge parameter valued in the (A-)dS$_2$ algebra.

The theory \eqref{JTBFaction} admits a second-order formulation. The proof works as follows: varying with respect to $\Phi^a$ yields $R^a(\tP)=0$, which allows solving for the spin connection in terms of the zweibein and its derivatives. On the other hand, varying with respect to $\Omega$ gives the equation $d\Phi-\tLambda\epsilon_{ab}E^a \Phi^b=0$, which allows expressing $\Phi^a$ in terms of $\Phi$. Plugging back $\Omega$ and and $\Phi^a$ into \eqref{JTBFaction}, the BF action takes the form of the JT gravity action. Since the fields $\{\Omega, \Phi^a\}$ were algebraically obtained from their own field equations, there is a dynamical equivalence between the first and the well-known second-order formulation of JT gravity, where $\Phi$ is the dilaton field (see, for instance, \cite{Pons:2009ch}). \\
 
A finite non-relativistic (NR) limit of JT gravity in the first-order formulation can be defined starting from a BF theory with gauge algebra (A-)dS$_2\times \mathbb{R} $ \cite{Gomis:2020wxp}. Thus we extend the (A-)dS$_2$ symmetry \eqref{adS2algebra.} by including an Abelian generator $\tY$, and its gauge field $X$. To carry out the contractions, we express the relativistic algebra generators with a linear combination of new generators that involves a dimensionless parameter $\varepsilon$. The NR contraction follows from defining NR one-form gauge fields $\tau$, $e$, $\omega$ and $m$, as
\bseq\label{NRfields}
\ba
\tau & = &  \varepsilon \, \left(E^0 +X \right) \,, ~~~~ ~~~e =  E^1\,, \\
m  &=& \frac{1}{\varepsilon} \left(E^0 -X \right)\,, ~~~~~~~\omega  =   \frac{1}{\varepsilon}   \Omega\,,
\ea
\eseq
 and also the definition of the NR scalar fields $\{\eta, \rho,\phi, \zeta\}$
 \bseq\label{NRscalars}
 \ba
 \eta & =& \frac{\varepsilon}{2} \, \left( \Phi^0 +\Psi \right) \,,~~~~~~\zeta  =  \frac{1}{\varepsilon} \,  \left( \Phi^0 - \Psi \right) \,,\\
 \phi & =&  \frac{\Phi}{\varepsilon} \,, ~~~~~~~~~~~~~~~~~~~~\rho  =   \Phi^{1}\,.
 \ea
 \eseq
 Replacing Eqs.~\eqref{NRfields} and Eqs.~\eqref{NRscalars} into the Eq.~\eqref{JTBFaction}, and defining the NR cosmological constant $\Lambda_{\scriptscriptstyle NR}$ and the constant $\gamma_0$ as
\be\label{rescLambda}
\Lambda_{\scriptscriptstyle NR}= \frac{1}{\varepsilon^2} \tLambda\,, \hspace{1.5cm} \gamma_0 = \varepsilon^2 \, \tilde{\gamma}_0\,,
\ee
we find, in the the limit $\varepsilon \rightarrow 0$, the NR two-dimensional gravity theory
\be \label{NRaction}
 S_{\scriptscriptstyle NR} = \gamma_0\int \Bigg(\phi R_{\scriptscriptstyle NR}(G) +\, \Lambda_{\scriptscriptstyle NR}\bigg(\eta R_{\scriptscriptstyle NR} (M) +\,\zeta R_{\scriptscriptstyle NR} (H)-\rho R_{\scriptscriptstyle NR} (P)\bigg)\Bigg)\,,
\ee
where
\bseq\label{NRcurvatures}
\ba
&&R_{\scriptscriptstyle NR}(H)  =  d\tau \,, ~~~~~~~~~~~~~~~~~~~~~~R_{\scriptscriptstyle NR}(P ) =  d e +\omega\tau\,,\\
&&R_{\scriptscriptstyle NR}(G)  =  d  \omega -\Lambda_{\scriptscriptstyle NR} \tau e\,, ~~~~~~~~~R_{\scriptscriptstyle NR}(M)  =   d m + \omega e\,.
\ea
\eseq
The action \eqref{NRaction} is invariant under the extended Newton-Hooke$_{\pm}$\footnote{
The plus sign corresponds to the contraction of dS algebra while the minus sign corresponds to the contraction of AdS algebra. Note that, the space-time symmetries of the harmonic oscillator form the NH$_2^{+}$ algebra, whereas the inverted harmonic oscillator has a NH$_2^{-}$ symmetry algebra.} algebra, given by
\be\label{extNewtonHooke}
[G,H] = P \,, ~~~~~~~[G,P] = M\,, ~~~~~~~[H,P] = -\Lambda_{\scriptscriptstyle NR}G\,,
\ee
where $H,P$ are generators of time and spatial translations, $G$ the Galilean boost generator, and $M$ is an additional abelian generator.  
It is worth mentioning that this action can be alternatively obtained as a BF theory based on the extended Newton-Hooke (NH) algebra \eqref{extNewtonHooke}, where the one-form gauge connection $A=A_\mu dx^\mu$ and the $B$ scalar field are given by
\be
\bal
A&= \tau H + e P + \omega G + m M\,,\\
B&= \eta H + \rho P + \phi G + \zeta M\,.
\eal
\ee
By considering the NR gauge symmetry parameter $\chi$ valued on the extended NH algebra as $\chi= \lambda H + \xi P + \Theta G + \Gamma M $, the symmetry transformations of the gauge fields are given by
\bseq\label{NRtrans}
\ba
\delta_{ \scriptscriptstyle NH}\tau_\mu& = & \partial_\mu \lambda \,, \\
\delta_{ \scriptscriptstyle NH} e_\mu& = & \partial_\mu \xi -\tau_\mu \Theta +\omega_\mu \lambda \,, \\
\delta_{ \scriptscriptstyle NH}\omega_\mu& = & \partial_\mu \Theta +\Lambda_{\scriptscriptstyle NR}(e_\mu \lambda- \tau_\mu \xi) \,, \\
\delta_{ \scriptscriptstyle NH} m_\mu& = & \partial_\mu \Gamma + \omega_\mu \xi - e_\mu \Theta \,, \\
\delta_{ \scriptscriptstyle NH} \eta & = &0\,,\\
\delta_{ \scriptscriptstyle NH}\rho & =& \lambda \phi-\Theta \eta \,,\\
\delta_{ \scriptscriptstyle NH}  \zeta & =& \xi \phi -\Theta \rho \,,\\
\delta_{ \scriptscriptstyle NH}\phi & = &\Lambda_{\scriptscriptstyle NR} (\lambda \rho- \xi \eta)\,.
\ea
\eseq

Like the relativistic case, we can move on to the second-order formulation of \eqref{NRaction}. Indeed, the field equations follow upon varying with $\rho$ and $\eta$, $R_{\scriptscriptstyle NR}(P)=0$ and $R_{\scriptscriptstyle NR}(M)=0$, allows to solve the NR spin-connection, we get
\be\label{NRomega}
 \omega_\mu =  2\, \tau^{[\alpha}e^{\beta ]} \,\left(  e_\mu \partial_\alpha e_\beta- \tau_\mu \partial_\alpha m_\beta  \right)\,.
\ee
Then, plugging \eqref{NRomega} into \eqref{NRaction}, we get the {\it magnetic extended NH gravity theory}
\ba \label{NRaction0lima}
S_{\scriptscriptstyle Mag. Ext. NH} =  \kappa \int d^2 x \det{(\tau e)}\, \phi \,\Bigg( \mathcal R_{\scriptscriptstyle NR}
-2\Lambda_{\scriptscriptstyle NR}\Bigg)\,,
\ea
where $\kappa= -\gamma_0/2 , \,\det{(\tau e)} \equiv - \epsilon^{\mu \nu} \tau_\mu e_\nu$, and 
\be
\mathcal R_{\scriptscriptstyle NR}=
8\tau^{[\mu} e^{\nu]}\partial_{\mu}\left(\tau^{[\alpha} e^{\beta]}\left(e_{\nu}\partial_{\alpha}e_{\beta}-\tau_{\nu}\partial_{\alpha}m_{\beta} \right)\right)\,.
\ee
Note that theories described by \eqref{NRaction} and \eqref{NRaction0lima} are not necessarily dynamically equivalent because the equations solved by  
\eqref{NRomega} are not those obtained by varying with respect to $\omega_\mu$. At any rate, in the sector $d\tau=0$, the two systems have equivalent field equations and the invariance of \eqref{NRaction0lima} under Newton-Hooke boost symmetry can be confirmed in the sector $d \tau=0$.\footnote{This is similar to the case of first-order gravity in four dimensions
$S=\kappa\int E\wedge E\wedge R(\Omega)$
which is invariant under local translations only when the torsion vanishes.} The action \eqref{NRaction0lima} corresponds to the NR second-order formulation of JT gravity \cite{Gomis:2020wxp}. \\

Another singular limit that can be carried out on the relativistic theory \eqref{NRaction} is the Carroll limit. This contraction is obtained from the observation that the Carroll (A-)dS$_2$ algebra admits a central extension in the commutator of  the Galilean boost and momentum 
generator,\footnote{The existence 
of this extension is a unique feature of the two-dimensional case since, unlike the Galilean case, the Carroll algebra does not admit a non-trivial 
central extension in four dimensions.}  
we name  this algebra {\it extended Carroll (A-)dS$_2$ algebra}.
One can see that this symmetry follows from the extended NH$_2^{\pm}$ algebra by interchanging the generators $H$ and $P$ and changing the sign of the cosmological constant $\Lambda_{\scriptscriptstyle NR}$.
This fact allows one to pass from Newton-Hooke to Carroll symmetries. The isomorphism reads as
\be
\bal
&{\rm Extended\;\,NH}_2^+\leftrightarrow{\rm Extended\;\, Carroll\; AdS}_2\\
&{\rm Extended\;\, NH}_2^- \leftrightarrow {\rm Extended \;\,Carroll \;dS}_2\,.
\eal
\ee
Then, the extended (A-)dS$_2$ Carroll algebra is defined as 
\be\label{extAdSCarroll}
[G,P] = H \,, ~~~~~~~[G,H] = M\,, ~~~~~~~[H,P] = -\Lambda_{\scriptscriptstyle C}G\,, 
\ee
where $\Lambda_{\scriptscriptstyle C}=-\Lambda_{\scriptscriptstyle NR}$. The gravity theories invariant under these two algebras are classically related through the interchange of its fields:  $\tau \leftrightarrow e$ and $\eta \leftrightarrow \rho$. Hence, the Carroll action can be obtained from the NR one \eqref{NRaction}, yielding 
\be \label{Carrollaction}
 S_{\scriptscriptstyle C}  = \gamma_0 \,  \int \Bigg(\phi R_{\scriptscriptstyle C}(G) + \Lambda_{\scriptscriptstyle C}\bigg(\rho R_{\scriptscriptstyle C} (M) +\,\zeta R_{\scriptscriptstyle C} (P)-\eta R_{\scriptscriptstyle C} (H)\bigg)\Bigg) \,,
\ee
with the Carroll curvature two-forms given by
\ba\label{Carrollcurvatures}
R_{\scriptscriptstyle C}(H) & = & d\tau+\omega e  \,,~~~~~~~~~~R_{\scriptscriptstyle C}(P ) =  d e\,,\\
R_{\scriptscriptstyle C}(G)  &=&  d  \omega -\Lambda_{\scriptscriptstyle C}  \tau e\,, ~~~~~~R_{\scriptscriptstyle C}(M)  =   d m +\omega  \tau\,.
\ea
The action \eqref{Carrollaction} is invariant under infinitesimal Carroll gauge transformations
\bseq\label{Carrolltrans}
\ba
\delta_{ \scriptscriptstyle C}\tau_\mu& = & \partial_\mu \lambda -\Theta e_\mu +\omega_\mu \xi, \\
\delta_{ \scriptscriptstyle C} e_\mu& = & \partial_\mu \xi \,, \\
\delta_{ \scriptscriptstyle C}\omega_\mu& = & \partial_\mu \Theta +\Lambda_{\scriptscriptstyle C}(e_\mu \lambda-\tau_\mu \xi) \,, \\
\delta_{ \scriptscriptstyle C} m_\mu& = & \partial_\mu \Gamma + \omega_\mu \lambda - \tau_\mu \Theta \,, \\
\delta_{ \scriptscriptstyle C} \eta & = & \xi \phi-\Theta \rho \,,\\
\delta_{ \scriptscriptstyle C} \rho & =& 0 \,,\\
\delta_{ \scriptscriptstyle C} \zeta & =& \lambda \phi-\Theta \eta \,,\\
\delta_{ \scriptscriptstyle C}\phi & = & \Lambda_{\scriptscriptstyle C}(\lambda \rho-\xi \eta) \,.
\ea
\eseq
The action \eqref{Carrollaction} admits a second-order formulation in the sector $R_{\scriptscriptstyle C}(H)=de=0$. We refer to it as the {\it electric extended Carroll gravity theory}
\be \label{Carroll2order}
S_{\scriptscriptstyle Elec. Ext. C} =   \kappa  \int d^2 x \det{(\tau e)}\, \phi \,\Bigg( \mathcal R_{\scriptscriptstyle C}
+2\Lambda_{\scriptscriptstyle C}\Bigg)\,,
\ee
where we have defined
\be
 \mathcal{R}_{\scriptscriptstyle C}  = 8 e^{[\mu} \, \tau^{\nu]} \, \partial_\mu \left(e^{[\alpha} \tau^{\beta]} ( \tau_\nu\partial_\alpha \tau_\beta  - e_\nu\partial_\alpha m_\beta ) \right)\,.
\ee
Unlike the relativistic case, we emphasize that the NR second-order actions \eqref{NRaction0lima} and \eqref{Carroll2order} are not dynamically equivalent to their first-order versions. This is because the equivalence holds only in the $d\tau =0$ and $de=0$ sectors.
\section{Electric and magnetic (extended) versions of non-relativistic and Carroll JT gravity}
Recent results in non-Lorentzian field theories reveal that one can derive two different NR/Carroll field theories invariant under the same symmetry group \cite{Henneaux:2021yzg,Bergshoeff:2022eog}. Following the ideas for the flat field theories of \cite{axelericq}, we show a way to obtain the so-called {\it electric/magnetic gravity theories} for NR and Carroll JT gravity. By considering the second-order formalism, the derivation of these gravity theories suggests the inclusion of Lagrange multipliers to constrain the geometry and matter.  

\subsection{Electric extended Newton-Hooke gravity theory}
Let us start with the extended Carroll version of second-order JT gravity
\eqref{Carroll2order}. We will now show how this Carroll gravity theory, which is invariant in the torsionless sector, can be modified to contain NH symmetry. The corresponding is given by
\be \label{ElecExtNH}
 S_{\scriptscriptstyle Elec.Ext.NH} = \kappa \, \int \, d^2x \Bigg( \det(\tau e)\,  \phi \, \left(\mathcal{R}_{\scriptscriptstyle C} +2\Lambda_{\scriptscriptstyle C} \right) +~\partial_\mu  \pi^{[\mu \nu]}  a_\nu\Bigg) \,,
\ee
where $\pi^{[\mu \nu]}\equiv 4\, \det(\tau e) \, \phi\,  e^{[\mu} \tau^{\nu]}$.  We call \eqref{ElecExtNH} the {\it electric extended Newton-Hooke gravity theory}. The variation of the action \eqref{Carroll2order} under NH boosts yields a term proportional to an NH invariant piece. Due to this, one can also apply the procedure proposed in \cite{axelericq}. Indeed, by using the NH gauge transformations for the vector fields
\bseq \label{transfupvectors}
\ba
\delta_{ \scriptscriptstyle NH}\tau^\mu & = & e^\mu \Theta-\left( e^\mu  \partial_\nu \xi+e^\mu  \omega_\nu \lambda+\tau^\mu  \partial_\nu \lambda\right) \tau^\nu  \\
\delta_{ \scriptscriptstyle NH}e^\mu & = & -\left(e^\mu  \partial_\nu \xi +e^\mu  \omega_\nu \lambda+\tau^\mu  \partial_\nu \lambda\right) e^\nu\,,
\ea
\eseq
one can easily show that the term $\pi^{[\mu \nu]}$ is an extended NH invariant. It is manifest that the field $a_\mu$ acts as a Lagrange multiplier enforcing the following conditions for the zweibein $e$, the clock one-form $\tau$ and the scalar field $\phi$
\be\label{constraintwithm}
\partial_\mu  \pi^{[\mu \nu]} =0 \,,
\ee
which are obtained varying \eqref{ElecExtNH} with respect to $a_\nu$.
One can verify that the last NR gravity theory \eqref{ElecExtNH} is invariant under boosts of the extended NH algebra if the auxiliary field $a_\mu$ transforms as
\be
\delta_{\scriptscriptstyle NH} a_\mu=2e^{[\alpha} \tau^{\beta]} \left(\partial_\alpha m_\beta \tau_\mu+ \partial_\alpha e_\beta e_\mu\right)\Theta-\tau^{\alpha}\partial_\alpha \Theta e_\mu\,.
\ee

It is also possible to modify \eqref{NRaction0lima} to be Carroll invariant by including Lagrange multipliers. In this way, one arrives at
\be \label{MagExtC}
 S_{\scriptscriptstyle Mag.Ext.C} = \kappa \, \int \, d^2x \Bigg( \det(\tau e)\,  \phi \, \left(\mathcal{R}_{\scriptscriptstyle NR} -2\Lambda_{\scriptscriptstyle NR} \right)  -\,\partial_\mu  \pi^{[\mu \nu]}  \, b_\nu\Bigg) \,,
\ee
where we have introduced the Lagrange multiplier $b_\mu$ transforming under boosts of the extended Carroll algebra as 
\be
\delta_{\scriptscriptstyle C} b_\mu=2\tau^{[\alpha} e^{\beta]} \left(\partial_\alpha m_\beta e_\mu+ \partial_\alpha \tau_\beta \tau_\mu\right)\Theta-e^{\alpha}\partial_\alpha \Theta \tau_\mu\,.
\ee
We call \eqref{MagExtC} the {\it magnetic extended Carroll gravity theory}. Note that using the isomorphism between extended NH and (A-)dS Carroll algebras, one can easily obtain \eqref{MagExtC} from \eqref{ElecExtNH} by interchanging $\tau \leftrightarrow e$. In the same way, the gauge transformations of the Lagrange multipliers $a_\mu$ and $b_\mu$ are related through the same mapping. This new action is invariant under the extended Carroll algebra.

\section{Electric and magnetic non-relativistic and Carroll JT gravity}
\subsection{NH Electric version}
In $(1+1)$-spacetime dimensions, the Carroll algebra without a central extension admits a degenerate invariant bilinear form $\langle P , P \rangle= \gamma_0\Lambda_{\scriptscriptstyle C}$. The BF theory that one can built from the two-dimensional Carroll algebra is
\be\label{2Dcarrollaction}
S_{\scriptscriptstyle C}[\rho,e]=\int \langle B,F\rangle=\gamma_0 \Lambda_{\scriptscriptstyle C} \int \rho \,R_{\scriptscriptstyle C} (P)\,,
\ee
with $B=\eta H + \rho P+ \phi G, F= dA+A^2, A= \tau H + eP+ \omega G$, with $H,P, G$ satisfying the commutation relations \eqref{extAdSCarroll}. As in the previous section, we can build an action invariant under the complete NH algebra from \eqref{2Dcarrollaction}. Demanding this condition, it is necessary to introduce specific auxiliary fields ensuring that the last action satisfies this requirement. This action takes the form 
\be \label{CarrollwLagrange}
S_{\scriptscriptstyle NH}= S_{\scriptscriptstyle C}\,+\, \gamma_0 \int \Bigg(\phi R_{\scriptscriptstyle NR} (G) +\,\Lambda_{\scriptscriptstyle NR}\Bigg(  \alpha \eta+\beta R_{\scriptscriptstyle NR}(H) - \rho \omega \tau \Bigg) \Bigg) \,,
\ee
where the NR curvatures $R_{\scriptscriptstyle NR} (H), R_{\scriptscriptstyle NR} (P)$, and $R_{\scriptscriptstyle NR} (G)$, are defined in Eq.~\eqref{NRcurvatures}. The two-form $\alpha$ and the scalar $\beta$ are new auxiliary fields. The invariance of this new action under the unextended NH algebra requires choosing $\Lambda_{\scriptscriptstyle C}=-\Lambda_{\scriptscriptstyle NR}$, and the following symmetry transformations for the auxiliary fields 
\ba \label{TransLaw}
\delta_{ \scriptscriptstyle NH}\alpha & = & \xi R_{\scriptscriptstyle NR} (G) - \Theta R_{\scriptscriptstyle NR} (P)\\
\delta_{ \scriptscriptstyle NH}\beta & = &  \xi\phi-\Theta \rho\,.\label{TransLaw2}
\ea
We stress that \eqref{TransLaw} and \eqref{TransLaw2} are not the transformation laws of a BF gauge theory. They imply that $\alpha$ and $\beta$ are not part of a gauge connection that allows the construction of a gauge theory.\\
Unlike in the previous section, for Carroll gravities without the field $m$, we cannot solve algebraically for $\omega$ in terms of the remaining fields and their derivatives since the field equations of \eqref{CarrollwLagrange} do not fix a specific component of $\omega$. Then, we can solve partially and analytically the spin connection $\omega$ from the field equation $R_{\scriptscriptstyle NR} (P)=0$ by projecting along the $\tau$ direction; we get 
\be
\omega_\mu = \omega_{\perp \mu}+ \omega_{\parallel \mu}\,,
\ee
where 
\be\label{spinconnectionNR}
\omega_{\perp \mu}=\left(2 \tau^{[\alpha} e^{\beta]} \partial_\alpha e_\beta \right)e_\mu \,,
\ee
which is clearly orthogonal to $\tau^\mu$ since $e_\mu \tau^\mu=0$. In the second order formulation, we will interpret the indeterminate part $\omega_{\parallel \mu}$ as a Lagrange multiplier \cite{Bergshoeff:2017btm}. In the torsionless sector $ d\tau=0$ with $\eta=0$, and integrating by parts,  we find the {\it electric NH gravity theory}
\be\label{magaction}
 S_{\scriptscriptstyle {\it Elec.NH} } = \mathring{S}_{\scriptscriptstyle C} \, + \, \kappa \int \hspace{-0.1cm}d^2x\,\partial_{ \mu}\pi^{[\mu \nu]} \,\omega_{\scriptscriptstyle\parallel \nu} \,,
\ee
with 
\be \label{CarElecAction}
\mathring{S}_{\scriptscriptstyle C}= \kappa \int d^2x \, \det (\tau e)\, \phi \left(\mathring{\mathcal{R}}_{\scriptscriptstyle C}~ +2\Lambda_{\scriptscriptstyle C} \right)\,,
\ee 
where we introduced the Carroll-invariant Ricci scalar $\mathring{\mathcal{R}}_{\scriptscriptstyle C}$, written in terms of $\omega_{\perp \mu}$ 
\be
\mathring{\mathcal{R}}_{\scriptscriptstyle C}:= 4\tau^{[\mu} e^{\nu]} \partial_{\mu} \omega_{\perp\nu} = 8\tau^{[\mu} e^{\nu]} \partial_{\mu}  \left( \tau^{[\alpha} e^{\beta]} \partial_\alpha e_\beta e_{\nu} \right)\,.
\ee
The action \eqref{magaction} written in this form allows for efficiently checking the invariance of each term. In fact, the first term $\mathring{S}_{\scriptscriptstyle C}$ is Carroll-boost invariant.
 We refer to \eqref{CarElecAction}  as the {\it electric Carroll gravity action}. This theory coincides with \eqref{NRaction0lima} for $m_\mu=0$ and identifying $\Lambda_{\scriptscriptstyle C} =-\Lambda_{\scriptscriptstyle NR} $. Therefore, there exists an equivalence between magnetic extended NH theory and electric Carroll theory when the  extra $U(1)$ field vanishes.\\
On the other hand, as in the previous section, the term $\pi^{[\mu \nu]}=4\, \det(\tau e) \, \phi\,  e^{[\mu} \tau^{\nu]}$ is Carroll-boost invariant. The invariance of $S_{\scriptscriptstyle {\it Elec. NH} } $ under the NH symmetries implies to that the indeterminate part $\omega_{\scriptscriptstyle\parallel } $ of the spin connection transforms as 
\be
\delta_{\scriptscriptstyle NH}\omega_{\parallel\mu}= e^\alpha e_\mu \partial_\alpha \Theta - 2\tau^{[\alpha} e^{\beta]} \left(e_\mu \partial_\alpha \tau_\beta +\tau_\mu \partial_\alpha e_\beta \right)\Theta\,.
\ee
We can interpret $\omega_{\scriptscriptstyle\parallel }$ as a Lagrange multiplier constraining the geometry and matter of the theory. The variation of the action $ S_{\scriptscriptstyle {\it Elec. NH} }$ with respect to $\omega_{\scriptscriptstyle\parallel }$ implies the constraint \eqref{constraintwithm} previously obtained. 
 
\subsection{Carroll magnetic version}
An analogue procedure can be followed to obtain the magnetic version of Carroll gravity theory via the mapping between NH and Carroll algebras. Starting with the BF theory based on the degenerate bilinear form $\langle H ,H \rangle= -\gamma_0\Lambda_{\scriptscriptstyle NR}$ of the NH algebra,  a new Carroll-invariant theory can be constructed. \\
 The {\it magnetic Carroll gravity theory}, obtained by interchanging $\tau$ and $e$ in \eqref{magaction} is given by
\be\label{unext.mag.Carroll}
S_{\scriptscriptstyle Mag.C}=  \mathring{S}_{\scriptscriptstyle NH}-\kappa \int \hspace{-0.1cm}d^2x\,\partial_{\scriptscriptstyle \mu}\pi^{[\mu \nu]} \,\omega_{\scriptscriptstyle\parallel \nu}\,,
\ee
where the {\it magnetic NH gravity theory} is defined as
\be 
\mathring{S}_{\scriptscriptstyle NH}= \kappa \int d^2x \, \det (\tau e)\, \phi \left(\mathring{\mathcal{R}}_{\scriptscriptstyle NR}~ -2\Lambda_{\scriptscriptstyle NR} \right)\,,
\ee
and the NH invariant Ricci scalar is
\be
\mathring{\mathcal{R}}_{\scriptscriptstyle NR}:= 8 e^{[\mu} \tau^{\nu]} \partial_{\mu}  \left(  e^{[\alpha} \tau^{\beta]} \partial_\alpha \tau_\beta \tau_{\nu} \right)\,.
\ee
The magnetic terminology for $\mathring{S}_{\scriptscriptstyle NH}$ is justified because it does not present extra fields. Again, we find a coincidence between the electric extended Carroll theory and magnetic NH theory for $m_\mu=0$ and $\Lambda_{\scriptscriptstyle NR}=- \Lambda_{\scriptscriptstyle C}$.
For this case, the invariance of \eqref{unext.mag.Carroll} under Carroll symmetry implies the following transformation for the indeterminate part of the spin connection (orthogonal to $\tau^\mu$)
\be
\delta_{\scriptscriptstyle C}\omega_{\parallel\mu}= \tau^\alpha \tau_\mu \partial_\alpha \Theta - 2e^{[\alpha} \tau^{\beta]} \left(\tau_\mu \partial_\alpha e_\beta +e_\mu \partial_\alpha \tau_\beta \right)\Theta\,.
\ee

Let us mention that the symmetry transformations of the gravity theories found here do not correspond to the ones found in the previous section imposing $m_\mu=0$. 

\section{Conclusions and Outlook}\label{outlook}
The mapping between non-relativistic and Carroll field theories \cite{axelericq} has been extended to the case of JT gravity. Here, we have shown that the second order
non-relativistic and Carroll JT gravity (depending on the vielbein
variables $\tau,e,$ and the $U(1)$ vector field $m$ introduced in \cite{Gomis:2020wxp}) can be deformed
with two more theories introducing Lagrange
multipliers that restrict the dynamics of the
previous theories. However, they have the same symmetry algebras; the only difference with the gravity theories introduced in 
\cite{Gomis:2020wxp}, is the transformation law for the Lagrange multipliers.
The basis of this construction is due to the isomorphism between NH and the
Carroll algebras.

It will be interesting to know the number of physical degrees of freedom of the eight theories we have constructed.

Our results suggest a deep connection between the physics of non-relativistic and Carroll gravities at least in $(1+1)$-dimensions. The extension of our results to the gravity theories in more dimensions is an interesting path to follow. The applications to other physical theories would also be useful.

\subsection*{Acknowledgments}
We acknowledge interesting discussions about the relation between Galilean and Carroll theories with Roberto Casalbuoni and Jose Figueroa-O'Farrill. We also thank Eric Bergshoeff and Axel Kleinschmidt for discussions in a related project. The work of J.Z and L.A was partially supported by Fondecyt grants 122086 and 3220805, respectively. The work of JG has been supported in part by MINECO FPA2016-76005-C2-1-P 
and PID2019-105614GB-C21 and from the State Agen\-cy for Research of the
Spanish Ministry of Science and Innovation through the Unit of Excellence
Maria de Maeztu 2020-203 award to the Institute of Cosmos Sciences
(CEX2019-000918-M). DH was supported by the Icelandic Research Fund grant 228952-051.

\appendix
\section{Symmetry of electric Newton-Hooke JT gravity}

In this appendix, we explicitly check the invariance of the action \eqref{magaction} under the extended Newton-Hooke boosts. To do this, we should note that $\mathring{S}_{\scriptscriptstyle C} $ can be rewritten, up to boundary terms, as
\ba
\n \mathring{S}_{\scriptscriptstyle C} & =& \kappa \int d^2x \, \det (\tau e)\, \phi \left(\mathring{\mathcal{R}}_{\scriptscriptstyle C}~ +2\Lambda_{\scriptscriptstyle C} \right)\\
\n &=&\kappa \int d^2x \, \det (\tau e)\, \phi \left(4\tau^{[\mu} e^{\nu]} \partial_{\mu} \omega_{\perp\nu}~ +2\Lambda_{\scriptscriptstyle C} \right)\\
&=&  \kappa \int d^2x \partial_\mu \pi^{[\mu \nu]}\omega_{\perp\nu}~ \, + \,2\Lambda_{\scriptscriptstyle C}\kappa \int d^2x \det (\tau e)\phi   \,.
\ea
Then, the invariance of $\mathring{S}_{\scriptscriptstyle C}$ is seen as follows: the second term in the last equation is invariant under NH boost symmetry ($\delta_{\scriptscriptstyle  NH}$) since $\delta_{\scriptscriptstyle NH} \phi = 0$ and $\delta_{\scriptscriptstyle NH} \det(\tau e) =0$. Also, from the transformation \eqref{transfupvectors} and $\delta_{\scriptscriptstyle NH} \phi = 0$ one can show  
\be
\delta_{\scriptscriptstyle NH} \left(\partial_\mu \pi^{[\mu \nu]}\right) = 0\,.
\ee
So, under NH boosts of $\mathring{S}_{\scriptscriptstyle C}$ transforms as
\be
\delta_{\scriptscriptstyle NH}\mathring{S}_{\scriptscriptstyle C}=  \kappa \int d^2x \,\partial_\mu \pi^{[\mu \nu]} \delta_{\scriptscriptstyle NH} \omega_{\perp\nu}\,.
\ee
Therefore, the invariance of the action \eqref{magaction} yields
\be
\n \delta_{\scriptscriptstyle NH} S_{\scriptscriptstyle {\it Elec.NH} } =  \kappa \int d^2x \,\partial_\mu \pi^{[\mu \nu]} \delta_{\scriptscriptstyle NH} \omega_{\perp\nu}  
 \,  + \, \kappa \int d^2x\,\partial_{\mu}\pi^{[\mu \nu]} \, \delta_{\scriptscriptstyle NH} \omega_{\parallel \nu} \,,
 \ee
which is satisfied if
\be
\delta_{\scriptscriptstyle NH} \omega_{\perp}  =- \delta_{\scriptscriptstyle NH} \omega_{\scriptscriptstyle\parallel }\,.
\ee
The same procedure can be applied to the other gravity theories.

\end{document}